\documentclass[oldversion]{aa}
\usepackage{graphicx}
\usepackage{txfonts}
\usepackage{natbib}

\newcommand{\be}{\begin{equation}}
\newcommand{\ee}{\end{equation}}

\begin{document} 
 
\title{The actual Rees--Sciama effect from the Local Universe}

\author{Matteo Maturi\inst{1,2}
  \and Klaus Dolag\inst{2}
  \and Andr\'e Waelkens\inst{2}
  \and Volker Springel\inst{2}
  \and Torsten En{\ss}lin\inst{2}
}

\titlerunning{The actual Rees--Sciama effect from the Local Universe}
\authorrunning{Matteo Maturi et al.}

\institute{Zentrum f\"ur Astronomie, ITA, Universit\"at 
   Heidelberg, Albert-\"Uberle-Str.~2, 69120 Heidelberg, Germany
   \and
   Max-Planck-Institute for Astrophysics, 
   Karl-Schwarzschild-Str. 1, D-85741 Garching, Germany
}

\date{\emph{Astronomy \& Astrophysics, submitted}} 

\abstract{
  Observations of the Cosmic Microwave Background (CMB) have revealed
  an unexpected quadrupole-octopole alignment along a preferred axis
  pointing toward the Virgo cluster. We here investigate whether this
  feature can be explained in the framework of the concordance model
  by secondary anisotropies produced by the non-linear evolution of
  the gravitational potential, the so-called Rees-Sciama (RS) effect.
  We focus on the effect caused by the local superclusters, which we
  calculate using a constrained high-resolution hydrodynamical
  simulation, based on the IRAS 1.2-Jy all-sky galaxy redshift survey,
  which reproduces the main structures of our Universe out to a
  distance of $110\,{\rm Mpc}$ from our Galaxy.  The resulting RS
  effect peaks at low multipoles and has a minimum/maximum amplitude
  of $-6.6\,\mu{\rm K}$/$ 1.9\,\mu{\rm K}$. Even though its quadrupole
  is well aligned with the one measured for the CMB, its amplitude is
  not sufficient to explain the observed magnitude of the
  quadrupole/octopole alignment.  In addition, we analyze the WMAP-3
  data with a linear matched filter in an attempt to determine an
  upper limit for the RS signal amplitude on large scales. We found
  that it is possible to infer a weak upper limit of $30\,\mu{\rm K}$
  for its maximum amplitude. } 

\maketitle

\section{Introduction}

The Wilkinson Microwave Anisotropy Probe (WMAP) observations of the
Cosmic Microwave Background (CMB) \citep{SP03.1,HI06.1} turned out to
be in very good agreement with the predictions of the concordance
$\Lambda$-CDM cosmology.  However, the WMAP data also present a number
of unexpected features at large angular scales. In particular, these
include the small amplitude of the quadrupole \citep{SP03.1}, as
already noticed to be present within the COBE data by \cite{HI96.1},
the alignment between the CMB quadrupole and octopole, which are seen
to point toward the direction of Virgo \citep{TE03.1}, the planarity
of the octopole \citep{DE04.1}, the alignment between the combined
quadrupole and octopole momenta with the equinox and the ecliptic
plane \citep{SC04.4, CO06.1}, the localized source of non-gaussianity
(the cold spot) on angular scale $\sim 10^{\circ}$ \citep{VI04.1,CR05.1},
and the asymmetry of the large-scale power between the two galactic
hemispheres \citep{ER04.1,HA04.1}. The latter seems to be the most
relevant one.

Different possible explanations for some of these features have been
proposed, including a small universe with a cut-off scale below the
cosmological horizon \citep{SP03.1}, anisotropic universes such as
Bianchi models \citep{BU96.1,KO97.1,JA05.1}, quintessence models
\citep{DE03.1} or local dust-filled voids \citep{IN06.1}.

In the framework of the concordance $\Lambda$-CDM cosmology, we
explore the possibility of explaining the above ``anomalies'' with the
secondary CMB anisotropies produced by the evolution of the
gravitational potentials of the local cosmological structures. These
anisotropies are usually split into the integrated Sachs-Wolfe (ISW)
effect \citep[ISW;][]{SA67.1,HU94.1} and into the Rees-Sciama effect
\citep[RS;][]{RE68.1, MA90.2, SE96.2}, which are produced by the
linear and the non-linear evolution of the potential, respectively. We
focus our attention on the RS effect although our simulation also
includes the local ISW effect. However, the former is dominant over
the latter in the volume we consider. We ignore distant structures
beyond $110\,{\rm Mpc}$, because their contribution to the RS effect
is negligible at low multipoles \citep[see for e.g.][]{SE96.2,CO02.2}.

\begin{figure*}[!t]
  \centering
    \includegraphics[angle=90,width=0.95\hsize]{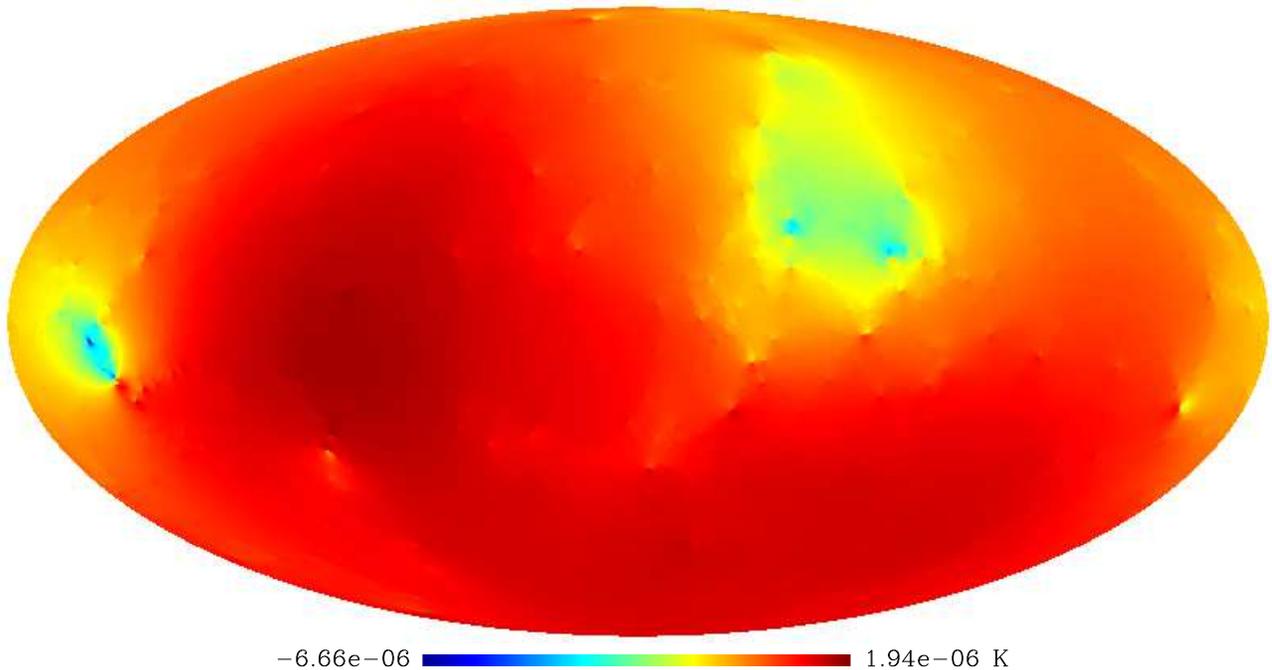}
    \caption{Full sky map of the Rees-Sciama effect produced by the
      local universe within a sphere of radius $110\,{\rm Mpc}$
      centered on our Galaxy. This CMB secondary anisotropy is
      dominated by the gravitational collapse of the most massive
      structures present in our constrained high-resolution
      hydrodynamical simulation. These are the galaxy clusters of
      Virgo, A3637, Centaurus, Hydra and Perseus. The maximum signal
      strength corresponds to an induced temperature shift of $-6.6\,
      \mu{\rm K}$ in the CMB.}
  \label{fig:RS_map}
\end{figure*}

A first study of the RS imprint from the local universe has been
carried out by \cite{CO05.1}. They used a toy-model with two spherical
mass concentrations of $8\times10^{15} h^{-1}M_\odot$ and $6.6\times
10^{15} h^{-1}M_\odot$ to represent the Great Attractor and the
Shapley concentration, and estimated a RS effect with a maximum
amplitude of $\sim 0.5 \mu{\rm K}$. Similarly, \cite{RA06.1} modeled
the Great Attractor with a spherical over-density
(Lama\^{i}tre-Tolman-Bondi model) with the Local Group falling into
it. They estimate a RS signal which may amount to $\Delta T/T \sim
10^{-5}$ and conclude that the RS effect is not compatible with the
features found in the WMAP data.

We use a high-resolution hydrodynamical simulation whose initial
conditions were constrained to reproduce the density and velocity
fields inferred with the IRAS 1.2-Jy all-sky galaxy redshift
survey. The simulation gives a realistic picture of the local universe
within a sphere of radius $110\,{\rm Mpc}$ centered on our Galaxy. We
use this simulation to obtain the expected RS effect on large scales
in the full sky, and to define an optimal linear filter for measuring
it in the WMAP-3 data.

The structure of the paper is as follows. In Section~2, we present the
numerical simulation and the computation of the full sky map of the RS
effect. In Section~3 we describe our analysis of the RS signal, and in
Section~4 we propose a method based on a matched optimal filter to
determine an upper limit for the amplitude of the RS effect in real
data. Finally, we present our conclusions in Section~5.

\section{Simulating the RS effect due to the local universe}

\subsection{The numerical simulation}

The results presented in this paper have been obtained by using the
$z=0$ output of a cosmological hydrodynamical simulation of the local
universe. We used initial conditions similar to those adopted by
\citet{MA02.1} in their study (based on a pure N-body simulation) of
structure formation in the local universe.  The galaxy distribution in
the IRAS 1.2-Jy galaxy survey is first smoothed with a Gaussian filter
on a scale of 7 Mpc and then linearly evolved back in time up to
$z=50$, following the method proposed by \cite{KO96.1}. The resulting
field is then used as a Gaussian constraint \citep{HO91.1} for an
otherwise random realization of a flat $\Lambda$-CDM model, for which
we assume a present matter density parameter $\Omega_{0}=0.3$, a
Hubble constant $H_0=70$ km/s/Mpc and r.m.s. density fluctuations of
$\sigma_8=0.9$ in spheres of comoving radius $8\,h^{-1}{\rm Mpc}$, with
$h=0.7$.

The volume that is constrained by the observational data covers a
sphere of radius $\sim 110\,{\rm Mpc}$ centered on the Milky Way. This
region is sampled with more than 50 million high-resolution dark
matter particles and is embedded in a periodic box of size $\sim 343$
Mpc on a side. The region outside the constrained volume is filled
with nearly 7 million low-resolution dark matter particles, allowing a
good coverage of long-range gravitational tidal forces that affect the
high resolution region.

The analysis by \citet{MA02.1} demonstrated that the evolved state of
these initial conditions provides a good match to the density and
velocity fields of the large-scale structure observed in the local
universe.  Unlike in the original simulation by \citet{MA02.1}, we
however also follow the evolution of the gas distribution. To
accomplish this we extended the initial conditions by splitting each
of the original high-resolution particles into a gas and a dark matter
particle, having masses of $0.48 \times 10^9\; M_\odot$ and $3.1
\times 10^9\; M_\odot$, respectively; this corresponds to a
cosmological baryon fraction of 13 per cent. The total number of
particles within the simulation is then slightly more than 108
million.

Our simulation run has been carried out with {\small GADGET-2}
\citep{SP05.1}, a new version of the parallel Tree-SPH simulation code
{\small GADGET} \citep{SP01.1} with an entropy-conserving formulation
of SPH \citep{SP02.1}.  The simulation neglected radiative cooling
processes and employed a comoving gravitational force resolution
(i.e.~the comoving softening length) of 14 kpc (Plummer-equivalent),
which is comparable to the inter-particle separation reached by the
SPH particles in the dense centers of our simulated galaxy clusters.
Previously, the same simulation has been used to study the propagation
of cosmic rays and to predict the SZ-effect from diffuse hot gas in
the local universe \citet{DO05.2,DO05.1}.

In this work we neglect the contribution of all structures placed at a
distance larger than $110\,{\rm Mpc}$, because the volume covered by
our simulation is sufficiently large to account for all relevant
sources of the RS effect that contribute to the {\em low} multipoles
we are interested in \citep[see for e.g.][where the total RS power
spectrum is shown to peak at $30<l$ ]{SE96.2,CO02.2}. For example, the
RS imprint of a galaxy cluster placed just outside the considered
volume would subtend RS features with scales typically smaller than
$\sim 0.5^{\circ}$, i.e. corresponding to multipoles larger than
$l\sim 650$ \citep[see for e.g.][]{MAT05.1}. We note that the RS
effect of galaxy clusters can also be related to their gravitational
lensing deflection field \citep{BI83.1} whose typical scale is at most
of the order of $1\,{\rm Mpc}$. The inclusion of the neglected large
volume beyond $110\,{\rm Mpc}$ would be possible through larger
numerical simulations \citep[see for e.g.][]{SP05.2} or through
analytical modeling. However, a direct simulation approach would in
general fail to reproduce the real matter distribution and dynamics,
and hence fail to match with its induced RS effect the actual
orientation of the low multipoles of the CMB.

\subsection{RS full sky map}\label{sec:RS_map}

Photons change their energy when they traverse evolving gravitational
potentials.  This effect introduces secondary anisotropies in the CMB,
which are usually split into two terms: the integrated Sachs-Wolfe
effect \citep[ISW,][]{SA67.1} and the Rees-Sciama effect
\citep[RS,][]{RE68.1}, produced by the linear and the non--linear
evolution of the potential, respectively. The resulting anisotropies
can be calculated as
\begin{equation}\label{eq:RS_integral}
  \frac{\Delta T}{T}(\vec\theta)= - \frac{2}{c^3}\int \dot{\Phi}
                                  (\vec{\theta}r,r/c)\, {\rm d}r \;,
\end{equation}
where $\dot{\Phi}$ refers to the time derivative of the gravitational
potential, $\vec\theta$ is the position on the sky, $r$ is the space
coordinate along the line-of-sight, and $r/c$ is the look-back time.
These secondary temperature anisotropies provide precious information
about the dynamical state and the evolution of the cosmic structures
between the last scattering surface and us. In particular, the signal
produced by the local universe could contribute to the low multipoles
of the observed CMB anisotropies in such a way to partially explain
the deviations from the expected Gaussian fluctuations as seen in the
WMAP data.

In order to estimate the secondary anisotropies due to the local
universe from our simulation we consider a Newtonian approximation
where the gravitational potential is defined as
\begin{equation}
  \Phi(\vec r,t) = -G \int {\rm d}^3r' \frac{\rho(\vec r',t)}{|\vec r - \vec r'|} \;,
\end{equation}
so that its time derivative is
\begin{equation}\label{eq:phidot}
  \dot\Phi(\vec r,t) = -G \int {\rm d}^3r' \rho(\vec r',t)\,\vec v(\vec r',t) 
  \frac{\vec r - \vec r'}{|\vec r - \vec r'|^3} \;,
\end{equation}
where $\rho$ and $\vec v$ are the matter density and velocity,
respectively.

We directly solve equation~(\ref{eq:phidot}) with a modified version
of the parallel code {\small GADGET-2} \citep{SP05.1}, whose original
version solves numerically a similar integral to compute the
gravitational forces. The RS map is finally obtained by adopting the
HEALPix pixelization of the sky \citep{GO05.2} with parameter $n_{\rm
  side}=512$, i.e. the pixels on the sky extend over $\sim 6.8'$ on a
side, and by evaluating equation~(\ref{eq:RS_integral}) at each pixel
position. We compute the integral along the line-of-sight as a direct
sum by sampling $\dot\Phi$ at regular intervals of $140\,h^{-1}{\rm
  kpc}$ between the radius $r_{\rm min}=1\,{\rm Mpc}$ and $r_{\rm
  max}=110\,{\rm Mpc}$. In doing so we include only the high
resolution region of the numerical simulation in order to avoid any
border effects. In this volume, the resolution is good enough to
properly sample the RS signal down to the typical cluster core scale.

The resulting full-sky RS map due to the local universe is shown in
Figure~\ref{fig:RS_map}. The main features are minima centered on
those structures of the numerical simulation which correspond to the
galaxy clusters of A3637, Virgo, Centaurus, Hydra and Perseus.  The
strongest of these signatures is related to Perseus (on the galactic
equator, left hand side) with a negative peak of $-6.6\,\mu{\rm K}$,
while the structure corresponding to the cluster of Coma shows a very
small contribution to the overall RS signal. The RS effects of other
prominent clusters are visible, but their overall impact is
negligible.  The RS map slightly favors negative values because we
sample a region of the universe populated by forming structures with
an ongoing gravitational collapse.

\section{Power spectrum analysis of the RS map}

\begin{figure}[!t]
  \centering
  \includegraphics[width=\hsize]{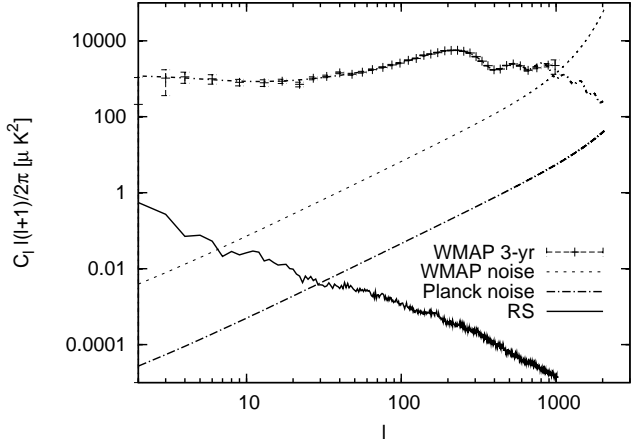}
  \caption{Angular power spectra of the CMB, the RS effect from the
    local universe and the noise of WMAP ($20\,\mu{\rm K}$ and ${\rm
    FWHM}=30'$) and of Planck ($10\,\mu{\rm K}$ and ${\rm
    FWHM}=5'$). The RS effect is well below the amplitude of the
    primary CMB, but for a few low multipoles it is larger than the
    instrumental noise level.}
  \label{fig:pwr_spectra}
\end{figure}

\begin{figure*}[!t]
  \centering
  \includegraphics[angle=90,width=8.5cm]{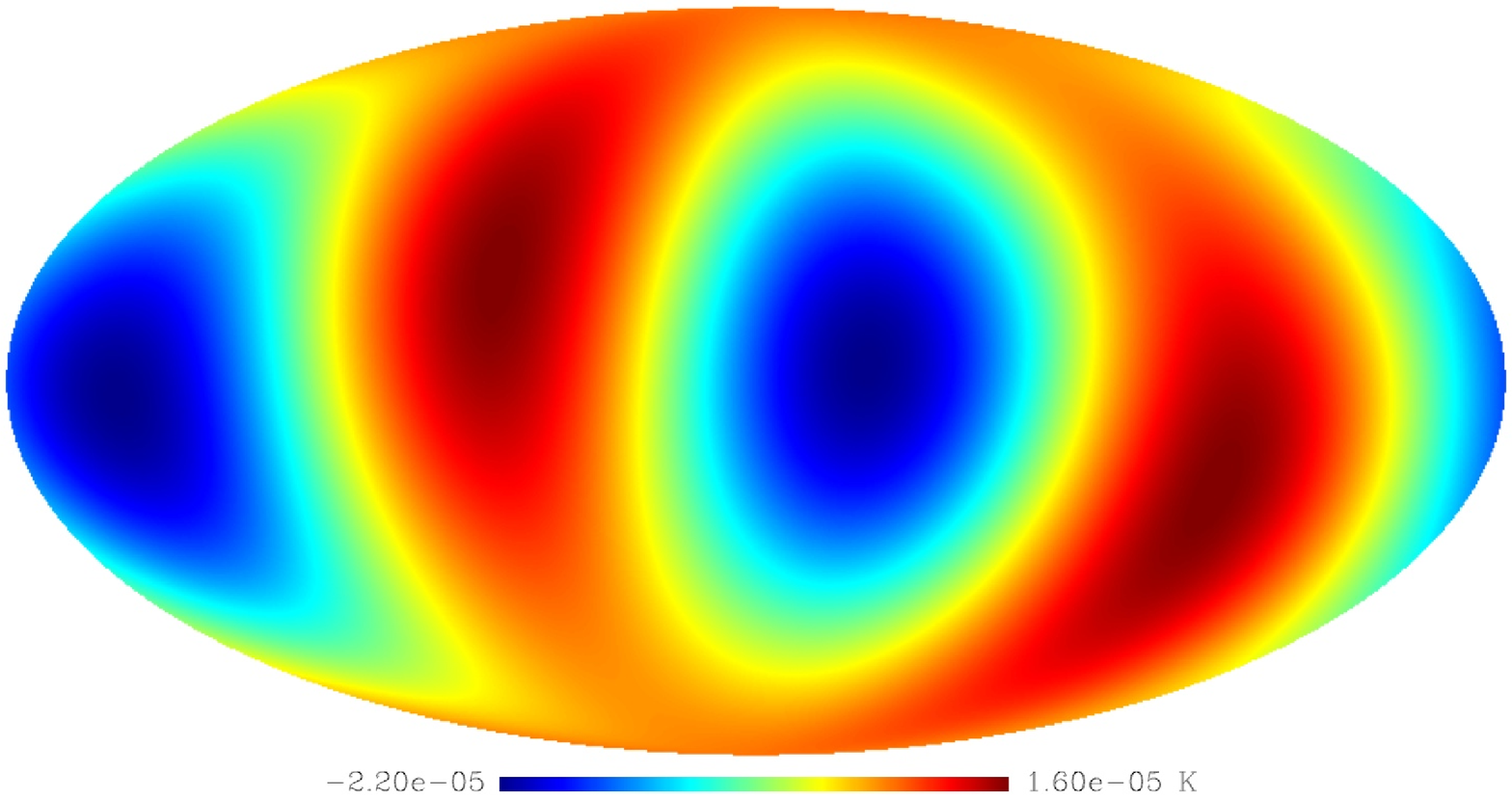}
  \includegraphics[angle=90,width=8.5cm]{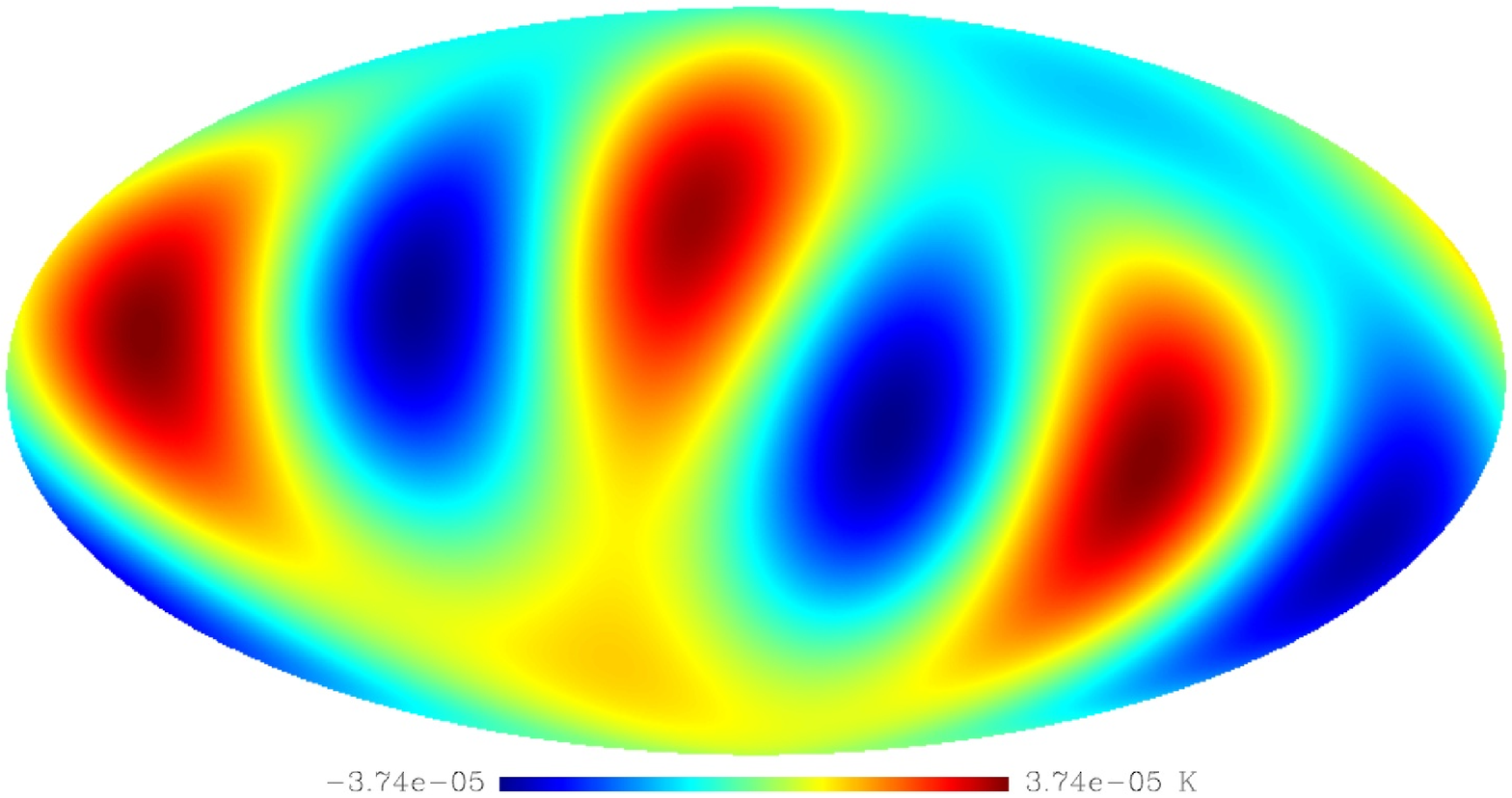}\\
  \includegraphics[angle=90,width=8.5cm]{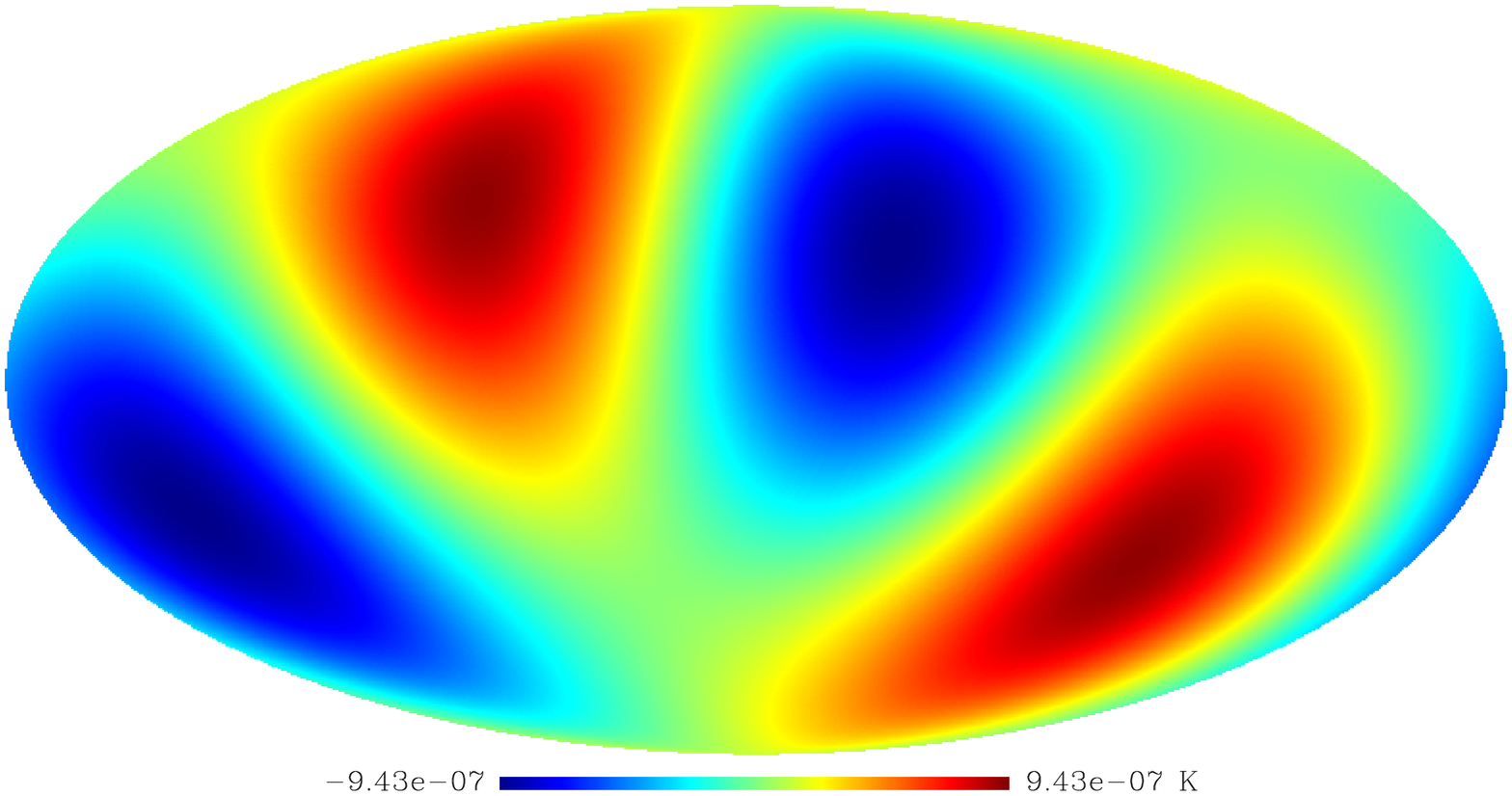}
  \includegraphics[angle=90,width=8.5cm]{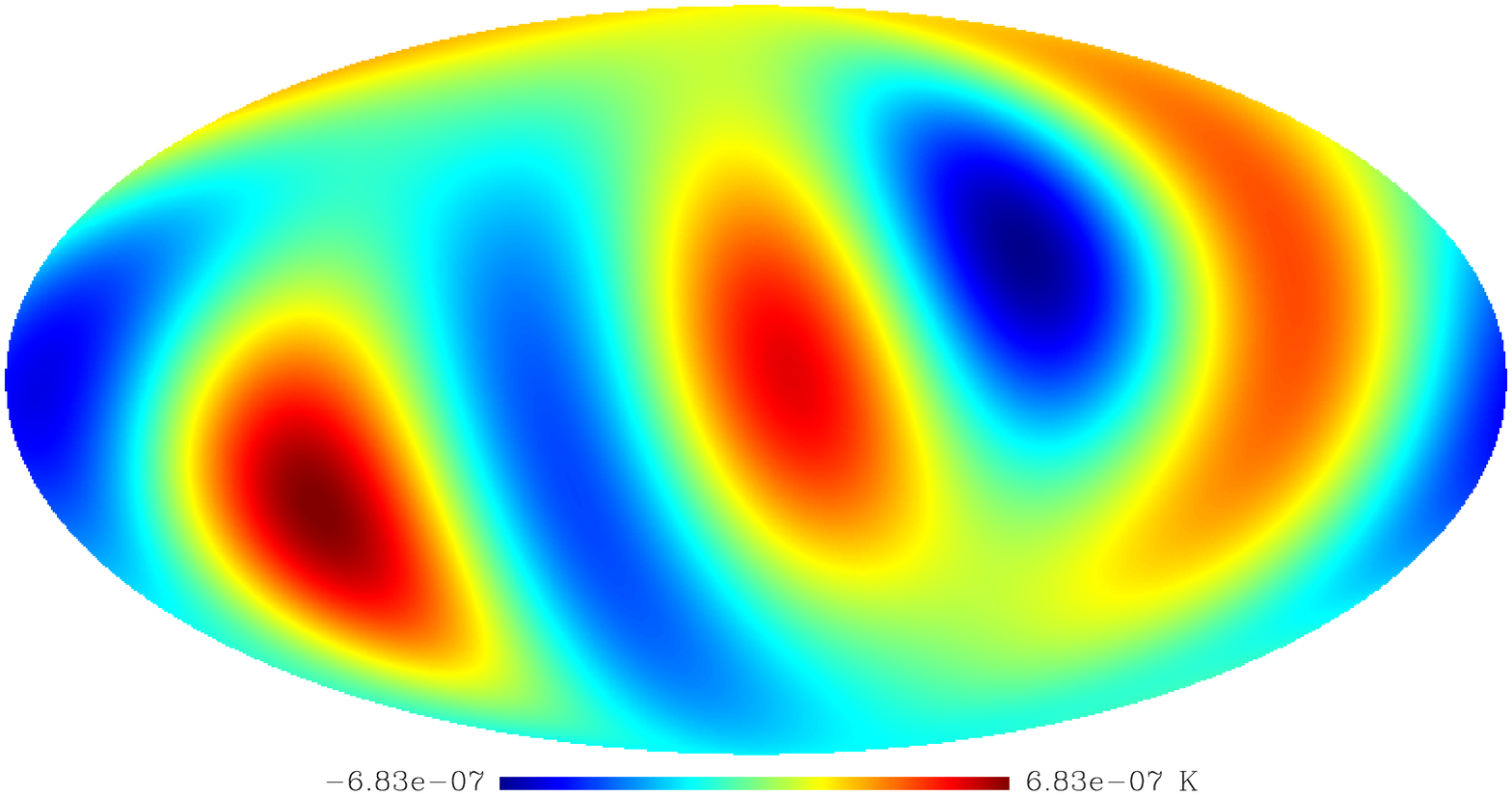}
  \caption{Comparison between the quadrupole and the octopole of the
    CMB as measured by WMAP (top panels) and of the RS effect from the
    local universe (bottom panels).}
  \label{fig:DQO_comparison}
\end{figure*}

The power spectra of our RS map, of the primary CMB, and of the
instrumental noise of WMAP and of the upcoming Planck experiment are
shown in Figure~\ref{fig:pwr_spectra}. The RS signal is dominated by
large-scale features which are only above the instrumental noise level
for a few multipoles ($l<6$ for WMAP and $l<30$ for Planck). However,
the RS signal is always well below the CMB primary fluctuations even
on the largest scales where our covered volume should be sufficient to
give the full expected signal. A detailed comparison between the
amplitudes of the dipole, quadrupole and octopole of the RS map and of
the CMB, as observed in the 3-years WMAP data \citep{HI06.1}, is given
in Table~\ref{tab:multiples}. It is curious to notice the similar
alignment of the quadrupole and octopole components of the RS and of
the CMB, see Figure~\ref{fig:DQO_comparison}. But in any case, the RS
effect from the local universe is far from being a relevant
contribution to the CMB anisotropies because of its small amplitude,
which corresponds only to a contribution at the percent level.

\begin{table}
\centering
\caption{Amplitude of the multipoles $l=1$, $l=2$ and $l=3$ of the RS effect
  as derived in our simulation, and of the CMB as measured by the 3 year WMAP
  data \citep{HI06.1}.}
\label{tab:multiples}
\begin{tabular}{|l|l|l|}
  \hline
  &
  WMAP                          & RS\\
  \hline
   l=1 dipole                    &
  removed                       & $\Delta T_{l=1} = 0.72 \,\mu{\rm K}$\\
  \hline
  l=2 quadrupole                &
  $\Delta T_{l=2}=14.5^{+15}_{-3}  \,\mu {\rm K}$ & $\Delta T_{l=2}
  = 0.74 \,\mu {\rm K} $\\
  \hline
  l=3 octopole                  &
  $\Delta T_{l=3}=32^{+26}_{-8} \,\mu {\rm K}$ & $\Delta T_{l=3}= 0.51
  \,\mu {\rm K}$\\
  \hline
\end{tabular}
\end{table}

We conclude that the RS effect due to the local universe does not
explain the deviations of the CMB from the expected primary
anisotropies predicted by the concordance $\Lambda$-CDM model,
i.e.~the non-Gaussian features and the quadrupole-octopole
alignment. However, we also found that the RS amplitude lies above the
instrumental noise level for low $l$, and thus it may be possible to
determine an upper limit for it, even if very weak, thanks to the
present or upcoming full sky CMB experiments.  On one hand a detection
would help to understand the local cosmological structures and the
details of their formation. On the other hand, if it was impossible to
achieve a detection with an optimally designed filter then this would
effectively ensure that the RS effect from the local universe
constitutes a negligible foreground for any kind of data analysis of
the CMB experiments. In the next Section, we therefore explore the
feasibility of measuring the RS signal from CMB data by using a linear
matched filter defined on the sphere.

\section{Measuring the RS effect}

We wish to construct a linear matched filter allowing the RS effect
from the local universe to be extracted from millimetric observations
of the CMB. We want the filter to suppress, as well as possible, the
primary anisotropies of the CMB as well as the instrumental noise. In
this work we ignore other possible sources of noise such us the
residuals of the galactic foreground separation. Because of the
significant level of such contaminations of the CMB primary
anisotropies we only expect to obtain an upper limit constraint.

\subsection{Non symmetric matched filter on the sphere}\label{sec:filter}

The filter used in this work is based on a template for the expected
signal and on the power spectrum of the noise. It was proposed by
\cite{HN96.1} in a flat sky approximation to estimate the peculiar
velocities of galaxy clusters through their kinetic
Sunyaev--Zel'dovich effect. Subsequently, it has also been proposed as
a tool to estimate the mass of galaxy clusters through weak lensing of
the CMB \citep{SE00.1} and for cluster detection in optical weak
lensing surveys \citep{MAT04.2}. An extension of the filter to the
sphere was first introduced by \cite{SA04.1}, assuming axially
symmetric models for the expected signal.  Since we are interested
only in the amplitude of the RS signal from the local universe and not
in recovering a map of the full sky, it is straightforward to
generalize the filter for non-axially symmetric models, like for the
RS effect. Our approach hence avoids the significant difficulties
in convolving the sky with a highly asymmetric filter.

We model the measured data $d(\vec{\theta})$, which contains a signal
$s(\vec\theta)=A\,\tau(\vec\theta)$ contaminated by some noise
$n(\vec\theta)$, by
\begin{equation} \label{eqn:signal} 
  d(\vec{\theta})=s(\vec\theta)+n(\vec\theta)= 
  A\tau(\vec\theta)+n(\vec\theta)\;.
\end{equation}

In our case, $n(\vec\theta)$ is mainly given by the sum of the CMB
primary anisotropies and the instrumental noise. Both can be
sufficiently well approximated with Gaussian random fields with zero
mean. The noise correlation function is
\begin{equation}
  \langle n^{}_{lm}\,n^*_{l'm'} \rangle = C_l\,\delta_{l,l'}
  \delta_{m,m'}\;,
\end{equation}  
where the asterisk denotes the complex conjugate, $\delta$ is the
delta function, and $C_l=C_{l,{\rm cmb}}+C_{l,{\rm inst}}$ is the
total angular power spectrum of the noise.

For convenience, all quantities are derived in the harmonic space.  We
wish to construct a linear filter $\Psi(\vec\theta)$ which yields an
estimate $\hat A$ for the amplitude $A$ of the signal, viz.
\begin{equation}\label{eq:estimate_A}
  \hat A = \sum_{l=0}^{+\infty} \sum_{m=-l}^{l} d_{lm} \Psi^*_{lm}\;,
\end{equation}  
where $d_{lm}$ and $\Psi_{lm}$ are the harmonic coefficients of the
data and of the filter function, respectively. The sum over $l$ is
usually restricted to the maximum multipole given by the resolution of
the observations.

The estimate is required to be unbiased and optimal, i.e.~the quantity
\begin{equation} 
  b \equiv\langle \hat A- A\rangle = A\left[ 
    \sum_{l=0}^{+\infty} \sum_{m=-l}^{l} \tau_{lm}\Psi^*_{lm}-1 
  \right] \;,
\end{equation} 
has to vanish and the r.m.s of the measurement
\begin{equation}\label{eq:variance}
  \sigma^2\equiv\left\langle(\hat A-A)^2\right\rangle= 
  b^2+ \sum_{l=0}^{+\infty} \sum_{m=-l}^{l} C_l \, \left|\Psi_{lm}\right|^2\;,
\end{equation} 
has to be minimal. The filter $\Psi$ which satisfies these two
conditions is the one which minimizes the action $L=\sigma^2+\lambda
b$, where $\lambda$ is a Lagrange multiplier. We thus obtain
\begin{equation} \label{eq:filter}
  \Psi_{lm}=\left[
    \sum_{l=0}^{+\infty} \sum_{m=-l}^{l}
    \frac{|\tau_{lm}|^2}{C_l} 
    \right]^{-1}\,
  \frac{\tau_{lm}}{C_l}\;.
\end{equation} 
This shows that the filter is most sensitive for those spatial
frequencies where the signal $\tau$ is large and the noise $C_l$ is
small. Note that this filter uses the full information available on
the expected signal $\tau$ and not only its power spectrum, as for
example a Wiener filter would do.

\subsection{An upper limit for the RS amplitude in the WMAP-3 data}

Before applying our procedure to real data, we have tested the
described filter on mock data where we combined the full sky RS map
(see Section~\ref{sec:RS_map}) from our simulation of the local
universe with a realization of the primary CMB fluctuations computed
with {\small CMBEASY} \citep{DO03.3} for a standard WMAP-3 cosmology.
The resulting map was then degraded by adding an instrumental noise
with the power spectrum
\begin{equation}\label{eq:Cl_instrument}
  C_{l,{\rm inst}}=w^{-1}\exp\left[
  \frac{l(l+1)\,\theta^2_{\rm FWHM}}{8\ln2}
  \right]\;,
\end{equation}
where $w^{-1}:=(\sigma_T \,\theta_{\rm FWHM})^2$ \citep{KN95.1}. The
resolution and sensitivity were fixed according to the characteristics
of the WMAP satellite $(\theta_{\rm FWHM}=30',\sigma_T = 20\, \mu{\rm
  K})$ and to the upcoming ESA space mission Planck ($\theta_{\rm
  FWHM}=5',\sigma_T = 10\, \mu {\rm K})$. A convolution with a
Gaussian kernel with the same FWHM is applied to the simulated
data. The dipole component of the signal is finally removed because of
the strong contamination which would be caused by the proper motion of
the solar system.  We ignore other contaminants such as residuals of
the foreground subtraction or extragalactic point sources.

The resulting map is finally analyzed with the filter described in
Section~{\ref{sec:filter}. For the filter template $\tau$ we use the
  same RS map which was also included in the simulated
  observations. This is a strong simplification, but here we only aim
  at testing the performance of the filter when all the conditions for
  its application are optimally satisfied.
  The results are as follows:\\

\begin{tabular}{lll}
Expected signal & : & $A= 1.9 \;\mu{\rm K}$\\
WMAP (simulation)   & : & $A= 3.4\pm  13\,\mu{\rm K}$ \\
Planck (simulation) & : & $A= 2.2\pm  12\,\mu{\rm K}$\\
\end{tabular}\\

\noindent
It is thus only possible to infer a weak upper limit for the amplitude
of the RS effect due to the local universe, because of the strong
contamination given by the CMB primary fluctuations.

Finally, we apply the filter, as previously described, to the 3-years
ILC map of the WMAP experiment \citep{HI06.1}. In this case, since the
ILC map is convolved with a Gaussian kernel with ${\rm FWHM}=1^\circ$,
we convolved the template $\tau$ in the same way before the filter
derivation. This convolution of the template has the implicit
advantage of suppressing small features in the RS template which may
not be present in the structures of the real universe. The resulting
filter is shown in Figure~\ref{fig:filter}.

To estimate the variance of the measurement, we refrain from using
Equation~(\ref{eq:variance}) because it supposes that the template is
identical to the expected signal and thus Eqn.~(\ref{eq:variance})
would yield a lower limit for the noise amplitude. We rather estimate
the noise level by applying the filter many times with the RS template
randomly rotated on the unit sphere. While the mean of these
randomized measurements should be consistent with zero because the
rotated templates do not match the expected signal, the dispersion
around zero provides a good estimate of the noise level. We obtain\\

\begin{tabular}{lll}
WMAP (ILC map) & : & $A=3 \pm 26\; \mu{\rm K}$ \\
\end{tabular}\\

which implies an upper limit for the amplitude of the RS effect in the
actual CMB data of $A_{\rm max}= 30\; \mu {\rm K}$. This is not very
restrictive compared to the expected signal strength of $A_{\rm exp} =
1.9\, \mu{\rm K}$.

\begin{figure}[!t]
  \centering
  \includegraphics[angle=90,width=8.5cm]{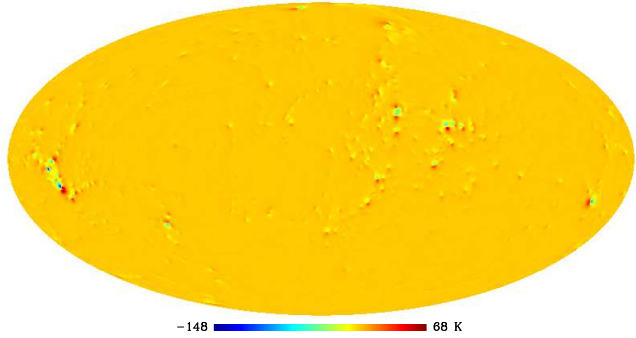}
  \caption{Filter function used to estimate the upper limit for the
    maximum amplitude of the RS effect produced by the local
    universe.}
  \label{fig:filter}
\end{figure}

\section{Conclusions}

We estimated a realistic map of the Rees-Sciama effect causd by the
evolution of the gravitational potential in the local universe within
a sphere with a radius of $110\,{\rm Mpc}$ centered on our Galaxy. To
this end we used a high-resolution hydrodynamical simulation whose
initial conditions were constrained to reproduce the density and
velocity fields inferred from the IRAS 1.2-Jy all-sky galaxy redshift
survey.

We estimated a minimum/maximum amplitude of the local RS effect of
$-6.6\,\mu{\rm K}/1.9\,\mu{\rm K}$, which is a factor of 10 larger
than the $0.5\mu{\rm K}$ predicted by \cite{CO05.1}. The power
spectrum of the RS signal lies above the instrumental noise of WMAP
and Planck for low multipoles but is always well below the CMB primary
anisotropies. Interestingly, the RS signal's quadrupole and octopole
moments are well aligned with those of the CMB primary
fluctuations. However, the small amplitudes of these RS multipoles are
not sufficient to explain the quadrupole-octopole alignment observed
by COBE and WMAP in the CMB.

In addition, we attempted to measure the RS effect on large scales in
mock CMB observations as well as in real data using a matched filter
technique.  We found that it is only possible to place a rather weak
upper limit for its amplitude even under optimistic assumptions.  In
analyzing the 3-years ILC map of WMAP \citep{HI06.1} this upper limit
turned out be $A_{\rm max}=30\, \mu {\rm K}$, which is not very
restrictive compared with the expected signal of $A_{\rm exp} = 1.9\,
\mu {\rm K}$.

\end{document}